\documentstyle[epsf]{article}
\begin{document}

\title{Computations of three-body continuum spectra}

\author{A. Cobis, D. V. Fedorov and A. S. Jensen\\
Institute of Physics and Astronomy,\\
Aarhus University, DK-8000 Aarhus C, Denmark}

\date{}
\maketitle

\begin{abstract}
We formulate a method to solve the coordinate space Faddeev equations
for positive energies. The method employs hyperspherical coordinates
and analytical expressions for the effective potentials at
large distances. Realistic computations of the parameters of the
resonances and the strength functions are carried out for the
Borromean halo nucleus $^6$He (n+n+$\alpha$) for J$^{\pi} = 0^{\pm},
1^{\pm}, 2^{\pm}$.
\end{abstract}

\paragraph*{Introduction.} The three-body continuum problem has been
the subject of numerous investigations \cite{glo96}. Tremendous
progress has been achieved, but still a number of problems remain
\cite{fri95}.  Many approximate solutions have been invented without an
emerging established general procedure.  Different treatments are
usually needed for short-range and long-range interactions and for
energies below or above possible two-body thresholds
\cite{car93,fri95a,kie96}. It is necessary, but not always easy, to
distinguish between inaccurate numerical results and shortcomings of
the basic interactions.

During the last decade a new class of weakly bound three-body systems,
nuclear halos, attracted enormous attention \cite{han95,fed94,zhu93}.
If no binary subsystem is bound, they are called Borromean nuclei.
These concepts are general and of interest in many subfields of
physics \cite{efi90,esr96}. Accumulating data from such systems demand
analyses heavily relying on the properties of their continuum spectra
\cite{dan93a,dan93b}. However, technical difficulties related to the
precise behavior at large distance are substantial and so far
unsolved.

Recently a new method with explicit analytical treatment of the large
distances \cite{fed93,fed96} was used to obtain bound-state solutions
to the Faddeev equations.  The method is very powerful as seen by the
successful investigation of the Efimov effect \cite{efi90,fed94a}.  The
purpose of this letter is to generalize the method to obtain continuum
state solutions.  In order to illustrate the efficiency of the method
we perform a realistic computation of a three-body Borromean halo
system.

\paragraph*{Method.} The $k$'th particle has mass $m_{k}$ and coordinate 
${\bf r}_k$. The two-body potentials are $V_{ij}$.  We shall use the
three sets of Jacobi coordinates (${\bf x}_i,{\bf y}_i$) and the
corresponding three sets of hyperspherical coordinates ($\rho$,
$\alpha_i$, $\Omega_{xi}$, $\Omega_{yi}$) \cite{fed94,zhu93,fed94a}.
The kinetic energy operator is then
\begin{eqnarray} \label{e10}
T=\frac{\hbar ^{2}}{2m}\left( -\rho ^{-5/2}\frac{\partial ^{2}}{\partial
\rho ^{2}}\rho ^{5/2}+\frac{15}{4\rho ^{2}}+\frac{\hat{\Lambda}^{2}}{\rho
^{2}}\right) \; , \\ \label{e15}
	\hat \Lambda^2=-{1 \over \sin(2\alpha)}
      {{\partial}^2\over {\partial} \alpha^2} \sin(2\alpha)  
	+{\hat l_x^2 \over {\sin^2 \alpha}}
 	+{\hat l_y^2 \over {\cos^2 \alpha}} -4  \; , 
\end{eqnarray}
where the angular momentum operators $\hat l_{x}^2$ and $\hat l_{y}^2$
are related to the ${\bf x}$ and ${\bf y}$ degrees of freedom.

The total wave function is now expanded in a complete set of
hyperangular functions
\begin{eqnarray} \label{e20}
\Psi (\rho ,\Omega )=
\frac{1}{\rho ^{5/2}}\sum_{n=1}^{\infty} f_{n}(\rho )
\sum_{i=1}^{3} \frac{\phi _{n}^{(i)}(\rho,\Omega_i)}{\sin(2\alpha_i)}
\; , \label{e23}
\end{eqnarray}
Each of the three components $\phi _{n}^{(i)}$ is expressed in the
corresponding system of Jacobi coordinates and they satisfy for each $\rho$
the three Faddeev equations
\begin{equation}\label{e25}
\big(\hat{\Lambda}^{2}-\lambda_n\big)
\frac{\phi_{n}^{(i)}}{\sin(2\alpha_i)}
+\frac{2m}{\hbar ^{2}}\rho ^{2}V_{jk} 
\sum_{l=1}^{3} \frac{\phi _{n}^{(l)}}{\sin(2\alpha_l)}
=0 ,\; 
\end{equation}
where $\{i,j,k\}$ is a permutation of $\{1,2,3\}$.  In the absence of
bound subsystems the eigenvalues $\lambda_n$ approach at large
distances the hyperspherical spectrum obtained for $V_{jk}=0$, i.e.
$\lambda_n(\rho \rightarrow \infty) = K_n(K_n+4)$, where $K_n$ is odd
or even natural numbers depending on the parity.

The expansion coefficients $f_{n}(\rho )$ satisfy the equations
\begin{eqnarray} \label{e30}
\left( -\frac{\partial ^{2}}{\partial \rho ^{2}}+\frac{\lambda _{n}+15/4}{%
\rho ^{2}}-Q_{nn}-\frac{2mE}{\hbar ^{2}}\right) f_{n}(\rho )
 \\ \nonumber
=\sum_{n^{\prime
}\neq n}\left( Q_{nn^{\prime }}+2P_{nn^{\prime }}\frac{\partial }{\partial
\rho }\right) f_{n^{\prime }}(\rho )\ .
\end{eqnarray} 

The coupling terms $P$ and $Q$ approach zero at least as fast as
$\rho^{-3}$. For Borromean systems we can then choose those
solutions $\Psi_{n'}$ to eq.(\ref{e20}) where the large-distance
($\rho \rightarrow \infty)$ boundary conditions for $f^{(n')}_{n}$ are
given by \cite{tay72}
\begin{equation} \label{e35}
f^{(n')}_n(\rho) \rightarrow \delta_{n,n'} 
 F^{(-)}_{n}(\kappa \rho) - S_{n,n'} F^{(+)}_{n}(\kappa \rho) \; ,
\end{equation}
where $\kappa^2=2mE/\hbar^2$ and $F^{(\pm)}_{n}$ are related to the
Hankel functions of integer order by
\begin{eqnarray}\label{e40}
F^{(\pm)}_{n}(\kappa \rho) = 
 \sqrt{\frac{m \rho}{4 \hbar^2}}\, H^{(\pm)}_{K_n+2}(\kappa \rho) \nonumber \\ 
 \rightarrow \sqrt{\frac{m}{2 \pi \kappa \hbar^2}} 
\exp\left[\pm i\kappa \rho \pm 
 {i\pi\over 2}(K_n+{3\over 2})\right] \; .
\end{eqnarray} 
The continuum wave functions are orthogonal and normalized to delta
functions in energy.

By diagonalization of the $S$-matrix we obtain eigenfunctions and
eigenphases. The phase shifts reveal the continuum structure of the
system. In particular, a rapid variation with energy indicates a
resonance.  A precise computation of resonances and related widths can
be done by use of the complex energy method, where eq.(\ref{e30}) is
solved for $E=E_r-i\Gamma/2$ with the boundary condition
$f^{(n')}_n=\delta_{n,n'} \sqrt{\frac{m \rho}{4 \hbar^2}}\,
H^{(+)}_{K_n+2}(\kappa \rho)$. These solutions correspond to poles of
the $S$-matrix \cite{tay72}.

\paragraph*{Large-distance behavior.} Eq.(\ref{e25}) can be solved for
large distances, where for short-range potentials all partial waves,
except s-waves, decouple.  We expand each component on the
hyperspherical basis with the quantum numbers
$\{l_x,l_y,L,s_x,s_y,S,J\}$ where L,S and J are the total orbital
angular momentum, total spin and total angular momentum, respectively.
We express two of the Faddeev components $(j,k)$ in the coordinates
related to the third Jacobi set $(i)$ and project out the partial wave
with a given set of angular momentum quantum numbers. This operation,
leading from the i'th to the j'th Jacobi coordinates, is denoted by
$R_{i,j}$.

For large $\rho$ only small $\alpha$ contribute to the terms
proportional to $V_{jk}(r_i)$ in eq.(\ref{e25}). This is due to the
assumption of short-range potentials and because $r_i \propto \rho
\sin \alpha_i$. Let us first explicitly consider the three coupled
components, $\phi^{(i)}_L$, characterized by $l_{xi}=0$ and $l_{yi}=L$
and therefore with the same total orbital angular momentum $L$ and
furthermore with the same spin structure. We expand in powers of
$\alpha_i$ and find the leading order contribution from the
transformation of such terms to be
\begin{eqnarray}\label{e45}
R_{i,j}\left[\frac{\phi^{(j)}_{L}(\rho,\alpha_{j})}
{\sin(2 \alpha_{j})}\right] \simeq
\frac{(-1)^L \phi^{(j)}_{L}(\rho,\varphi_{j,i})}
{\sin(2 \varphi_{j,i})} \; ,  \\  \label{e67}
 \tan \varphi_{i,j} = (-1)^p\sqrt{\frac{m_k(m_i+m_j+m_k)}{m_i m_j}} \; ,
\end{eqnarray}
where $p$ is the parity of the permutation $\{i,j,k\}$. Non-zero
$l_{xi}$-values had produced higher powers of $\alpha_i$ in
eq.(\ref{e45}). Thus, the eigenvalues $\lambda$ related to the other
partial waves decouple at large distances and approach the
hyperspherical spectrum. These waves assume the asymptotic behavior on
a distance scale defined by the range of the interactions. On the
other hand the s-waves couple and feel consequently the interactions
over a distance defined by the scattering lengths.

We shall now concentrate on a system consisting of two neutrons and a
spin-zero core.  This model directly applies to $^6$He, a halo nucleus
for which a large amount of experimental data exists.  The model is
also a good approximation for another halo nucleus, $^{11}$Li
\cite{zhu93}.

Due to the antisymmetry between neutrons the three coupled components
($l_{xi}=0$, $l_{yi}=L$, $i=1,2,3$) reduce to two and the angular
Faddeev equations eq.(\ref{e25}) are to leading order in $\alpha$
(large $\rho$) given by

\begin{eqnarray} \label{e50}
  \left(
 - \frac{\partial^2}{\partial \alpha_1^2} + \frac{L(L+1)}{\cos^2\alpha_1}
 + \rho^2 v_{\mbox{\scriptsize NN}}(\rho \sin{\alpha_1}) - \nu^2
   \right) \\  \nonumber 
 \times \phi_{L}^{(1)}(\rho,\alpha_1) = 
- 2 \alpha_1 (-1)^L \rho^2 v_{\mbox{\scriptsize NN}}
(\rho \sin{\alpha_1})  C^{(1)}_{L} \; , \\
  \left(
 - \frac{\partial^2}{\partial \alpha_2^2} + \frac{L(L+1)}{\cos^2\alpha_2}
 + \rho^2 v_{\mbox{\scriptsize N}c}(\rho \sin{\alpha_2}) - \nu^2
   \right) \label{e55}  \\  \nonumber 
 \times \phi_{L}^{(2)}(\rho,\alpha_2) =
- 2 \alpha_2 (-1)^L \rho^2 v_{\mbox{\scriptsize N}c}(\rho \sin{\alpha_2})  C^{(2)}_{L} \; ,
\end{eqnarray}
where $\nu^2 = \lambda +4$, $v_{\mbox{\scriptsize NN}}(x_1) =
 V_{23}(x_1/\mu_{23}) 2m/\hbar^2$, $v_{\mbox{\scriptsize N}c}(x_2) =
 V_{13}(x_2/\mu_{13}) 2m/\hbar^2$, $m\mu^2_{jk}=m_jm_k/(m_j+m_k)$,
\begin{equation}\label{e69}
   C^{(1)}_{L} = 2  \frac{\phi^{(2)}_{L}(\rho,\varphi)}{\sin(2 \varphi)} , 
 C^{(2)}_{L} =  \frac{\phi^{(1)}_{L}(\rho,\varphi)}{\sin(2 \varphi)}
 + \frac{\phi^{(2)}_{L}(\rho,\tilde{\varphi})}
{\sin(2 \tilde{\varphi})} \; .
\end{equation}
with $\varphi = \varphi_{12}$, $\tilde \varphi = \varphi_{23}$.

For large $\rho$ the short range potentials $\rho^2
v(\rho\sin{\alpha_i})$ vanish for all $\alpha_i$ except in a narrow
region around zero. Due to this rescaling the effective range
approximation becomes better with $\rho$ increasing and
therefore any potential with the same scattering length and effective
range would lead to the same results.  Let us then in the region of
large $\rho$ use square well potentials $V_{jk}(r) = - V^{(i)}_0
\Theta(r<R_i)$, or equivalently expressed by the reduced quantities
$v_{jk}(x) = - v^{(i)}_0 \Theta(x<X_i=R_i \mu_{jk})$, where the
parameters are adjusted to reproduce the given two-body scattering
lengths and effective ranges. The corresponding solutions are then
accurate approximations to our original problem at distances larger
than $2R_i$ \cite{jen97}.

The potentials $v(\rho \sin{\alpha_i})$ are zero when
$\alpha_i>\alpha^{(i)}_0 \equiv \arcsin(X_i/\rho)$. Then
eqs.(\ref{e50}-\ref{e55}) are especially simple, i.e.
\begin{equation} \label{e72}
  \left(
 - \frac{\partial^2}{\partial \alpha_i^2} + \frac{L(L+1)}{\cos^2\alpha_i}
 -\nu^2  
  \right) \phi_{L}^{(i)}(\rho,\alpha_i) = 0 \; 
\end{equation}
and the solutions, vanishing at $\alpha_i=\pi/2$, are given by
\begin{eqnarray}\label{e76}
 \phi_{L}^{(i,II)}(\rho,\alpha_i) = A_L^{(i)} 
 P_{L}(\nu,\alpha_i) \; , \label{e82}  \\
 P_{L}(\nu,\alpha) \equiv \cos^L\alpha
 \left(
 \frac{\partial}{\partial \alpha} \frac{1}{\cos\alpha}
\right)^L \sin\left[ \nu 
\left(\alpha - \frac{\pi}{2} \right) \right] \; .
\end{eqnarray}

The potentials $v(\rho \sin{\alpha_i})$ are for large $\rho$ only
finite when $\alpha_i < \alpha^{(i)}_0 \ll 1$. Then eqs.(\ref{e50})
and (\ref{e55}) are 
\begin{eqnarray}\label{e86}
  \left(
 - \frac{\partial^2}{\partial \alpha_i^2} 
+ L(L+1) - \rho^2 v_0^{(i)} -\nu^2  
  \right) \phi_{L}^{(i)}(\rho,\alpha_i) = \nonumber \\
 + 2 \alpha_i (-1)^L \rho^2 v_0^{(i)}  C^{(i)}_{L} \; ,
\end{eqnarray}
where the wave functions in $C^{(i)}_{L}$ in eq.(\ref{e69}) must be
$\phi^{(i,II)}_{L}$.  The solutions to eq.(\ref{e86}) are then
\begin{eqnarray} \label{e89}
\phi_{L}^{(i,I)}(\rho,\alpha) = B_L^{(i)}\sin(\kappa_i\alpha) 
- 2 \alpha (-1)^L \frac{\rho^2 v_0^{(i)}}{\kappa_i^2}  C^{(i)}_{L} 
   \; , \\  \label{e92}
\kappa_i^2 \equiv -[ L(L+1) - \rho^2 v_0^{(i)} - \nu^2 ]  \; .
\end{eqnarray}

Matching the solutions, eqs.(\ref{e76}) and (\ref{e89}), and their
derivatives at $\alpha_i=\alpha^{(i)}_0$ gives a linear set of
equations for $A_L^{(i)}$ and $B_L^{(i)}$. Physical solutions are then
only obtained when the corresponding determinant is zero.  This is the
quantization condition for $\lambda$ and the eigenvalue equation
determining the asymptotic behavior of $\lambda(\rho)$.

\paragraph*{Realistic computations for $^6$He.} The practical 
implementation of the me\-thod is tested on $^6$He considered as two
neutrons and a $^4$He-core.  The two-body interactions reproduce
accurately the s-, p- and d-phase shifts up to 20 MeV. Furthermore, a
diagonal three-body force, $S_3 \exp(-\rho^2 /b^2_3)$, is added
in eq.(\ref{e30}) for fine tuning.  The range of the three-body force
is by its definition given in terms of the hyperradius. For $^6$He,
$\rho$=2 fm and 3 fm correspond roughly to configurations where the
neutrons respectively are at the surface of the $\alpha-$particle and
outside the surface by an amount equal to their own radius.  The idea
of using the three-body force is to include effects beyond those
accounted for by the two-body interactions.

\begin{figure}[t]
\centerline{\epsfbox[100 100 300 300]{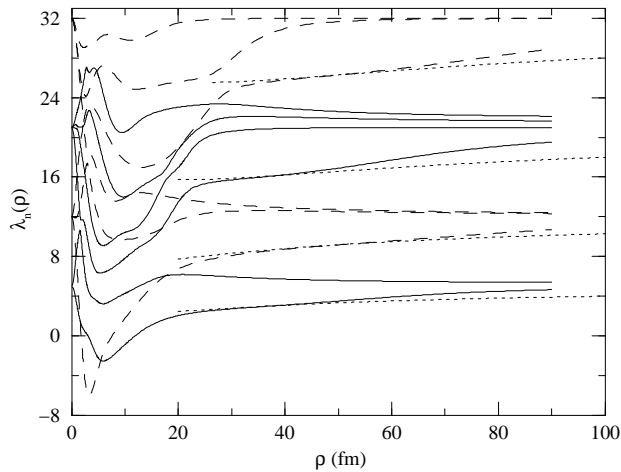}}
\caption{\small{
The lowest angular eigenvalues $\lambda_n$ as
function of $\rho$ for angular momentum $J^{\pi}=1^{-}$ (solid lines),
and $2^{+}$ (dashed lines) for $^6$He. The dotted lines are the
large-distance asymptotic behavior.  The neutron-neutron and the
neutron-$^4$He interactions are from \protect\cite{gar97} with
$(V_c^{(l)},r_c^{(l)})$ = (48.2 MeV, 2.33 fm), (-47.40 MeV, 2.30 fm),
(-21.93 MeV, 2.03 fm) for $s,p$ and $d$ waves respectively. The spin
orbit parameters are $(V_{so},r_{so})$ = (-25.49 MeV, 1.72 fm). Maximum
$K_n$-values up to 142 are used.
}}
\end{figure}

Several phase equivalent parametrizations are possible for each radial
shape of the two-body potential. They differ in the number of two-body
bound states of which the lowest s-state is occupied by the core
neutrons and therefore subsequently has to be excluded in the
computation. The results are very close after fine tuning by use of
the three-body interaction \cite{gar97}. We shall therefore only use
the potentials without bound states.  All possible s-, p- and d-waves
are included whereas other waves can be ignored to the accuracy we
need. The number of Jacobi polynomials in the basis expansion is
carefully chosen to give accurate numerical results up to a distance,
typically around 40 fm, where the asymptotic behavior is reached and
from then on the asymptotic solutions eqs.(\ref{e76}) and
eq.(\ref{e89}) are used.

The accurate low-energy continuum spectrum calculations require
integration of the radial equations up to distances of the order of ten
times the sum of the scattering lengths. For the n+n+$\alpha$ system
this is about 180 fm.  Too small basis size and too small maximum
distance are both disastrous for the numerical reliability.

In Fig. 1 we show the two (strictly decoupled ) angular eigenvalue
spectra for $1^{-}$ and $2^{+}$. The structure is complicated at small
distances where avoided level crossings are seen. The lowest level has
in both cases an attractive pocket unable to bind the system, but
still responsible for several resonances. At large distance the
structure is simpler as the hyperspherical spectrum is approached. In
the computation we use the asymptotic behavior, also shown on
Fig.1. This improvement of the procedure is absolutely essential when
accurate results are required.

\begin{figure}[t]
\centerline{\epsfbox[100 100 300 300]{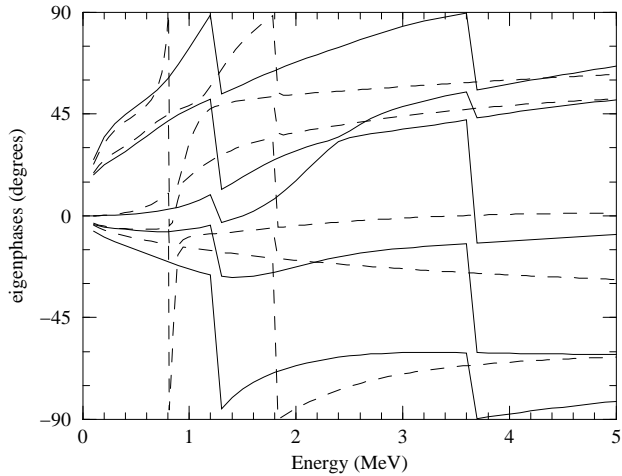}}
\caption{\small{
The eigenphases corresponding to the lowest
$\lambda$-values obtained after diagonalization of the $S$-matrix for
$J^{\pi}=1^{-}$ (solid lines) and $J^{\pi}=2^{+}$ (dashed lines). The
interactions are the same as in Fig.1 where a diagonal three-body
interaction, $S_3\exp(-\rho^2/b^2_3)$ with $S_3=-31$ MeV,
$b_3=2.061$ fm, is added in all partial waves.
}}
\end{figure}

The phase shifts for the cases in Fig. 1 are shown in Fig. 2.  The
rapid variation and the subsequent crossing of $\pi/2$, seen at four
energies, are the traditional signs of resonances. On the other hand it
is also possible to have resonance-like structures without phase shifts
crossing $\pi/2$.  They may still show up as poles of the S-matrix.

In Table 1 we give for a few spins and parities the two lowest
$S$-matrix poles obtained by the complex energy method.  The radial
equations were integrated numerically up to $\rho_{max}$=180 fm where
the $1/\rho^3$ tail of the effective potential becomes negligible.  The
numerical solutions were then matched at $\rho_{max}$ with the Hankel
functions $H^{(\pm)}$.  Precisely at the pole only the $H^{(+)}$
function must match the numerical solution.  We show the results for
two different three-body forces, i.e.~fine tuned to the ground state
energy and to the $2^{+}$-resonance.  Although the differences appear
to be relatively small, they are important for the observable
properties.  The two cases in Table 1 can be considered to give the
realistic range of the possible variation of the three-body force.  For
angular momentum $0^{-}, 1^{\pm}$ and $2^{-}$ the relatively small
pocket in the effective radial potential combined with strong
centrifugal barrier hinders the effect of the three-body force on the
phase shifts and the resonance properties.  Apart from the lowest 2$^+$
state all these resonances reside above the effective centrifugal barrier
and must therefore correspond to rather smooth structures in the
cross sections.

\begin{table}[t]
\caption{\small{The real and imaginary values $(E_r,\Gamma)$ (in MeV)
of the two lowest $S$-matrix poles $E=E_r-i\Gamma/2$ for $^6$He for
various spins and parities.  The interactions used are the same as in
Fig.1. The three-body interaction parameters are $S_3=-7.55$ MeV,
$b_3=2.9$ fm and $S_3=-31$ MeV, $b_3=2.061$ fm respectively for the
first two and the last two columns. Correspondingly the excitation
energies are $E^*=E_r + 0.95$ MeV and $E^*=E_r + 1.54$ MeV.}}
\begin{center}
\begin{tabular}{c|cc|cc||cc|cc|}
$J^{\pi}$ & $E_r$ & $\Gamma$ & $E_r$ & $\Gamma$ & $E_r$ & $\Gamma$ &
$E_r$ & $\Gamma$ \\ \hline 
$0^{+}$ & 0.94 & 0.64 & 1.46 & 0.83 & 0.62 & 0.56 & 1.16 & 0.67 \\
$0^{-}$ & 2.07 & 0.74 & - & - & 2.07 & 0.74 & - & - \\
$1^{+}$ & 1.62 & 0.74 & 2.55 & 0.86 & 1.62 & 0.74 & 2.55 & 0.86 \\
$1^{-}$ & 1.11 & 0.42 & 1.67 & 0.58 & 0.95 & 0.38 & 1.43 & 0.56 \\
$2^{+}$ & 1.02 & 0.37 & 1.23 & 0.45 & 0.845 & 0.093 & 1.05 & 0.40 \\
$2^{-}$ & 0.90 & 0.34 & 1.82 & 0.57 & 0.90 & 0.34 & 1.82 & 0.57 \\
\end{tabular} 
\end{center}
\label{tab1} 
\end{table}

\begin{figure}[t]
\centerline{\epsfbox[100 100 300 300]{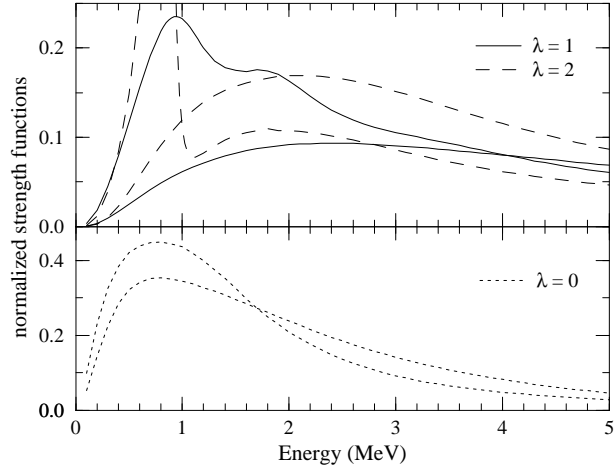}}
\caption{\small{
The strength functions, 
$S^{(\lambda)}(E) = \sum_{n} \left| \langle n J^{\pi} || M(E\lambda) ||
0^{+} \rangle \right|^2$ for $^6$He as function of energy for
transitions from the ground state to $0^{+}$ (dotted), $1^{-}$
(solid) and $2^{+}$ (dashed) excited continuum states.  The operator is
$M(E\lambda,\mu) = \rho^2$ and $\sum_{i=1}^3 q_i r_i^\lambda Y_{\lambda
\mu}(\hat r_i)$ respectively for $\lambda=0$ and $1,2$. The units are
the corresponding sum rules $\langle 0^{+}| \rho^4 | 0^{+} \rangle -
\langle 0^{+}| \rho^2 | 0^{+} \rangle^2$=749~fm$^4$ for $\lambda=0$
and $q_\alpha^2 (2\lambda+1) \langle 0^{+}|r_\alpha^{2\lambda} |0^{+}
\rangle /(4\pi)$=1.31$e^2$fm$^2$ and 8.19$e^2$fm$^4$ respectively for
$\lambda=1,2$, where $q_\alpha=2e$ is the $^4$He-charge and $r_\alpha$
is the $^4$He-distance from the $^6$He center of mass.  The
interactions are the same as in Fig.1 with a diagonal three-body
interaction added in all partial waves. The parameters are $S_3=-7.55$
MeV, $b_3=2.9$ fm for $0^{+}$ and $S_3=-31$ MeV, $b_3=2.601$ fm for
$1^{-}$ and $2^{+}$.  The smooth curves correspond to plane waves for
the continuum states.
}}
\end{figure}

An observable less sensitive to the large-distance behavior is related
to the excitations from the ground state. We show in Fig. 3 the lowest
three strength functions, $S^{\lambda}_{0^{+} \rightarrow J^{\pi}}(E)$,
as functions of energy both for plane waves and for the proper
continuum wave functions.  We find 91\%, 60\% and 70\% of the strength
below 5 MeV, respectively for monopole, dipole and quadrupole
excitations.  We notice the usual rise from zero to a maximum and the
fall off towards zero at large energy.  The peak is very pronounced for
$2^{+}$ reflecting the observed resonance of width 0.11 MeV at 0.82
MeV.  Above the smooth plane-wave background for $1^{-}$ is seen a peak
at about 0.95 MeV and a shoulder at about 1.8 MeV. This significant
$1^{-}$ enhancement is the result of a combination of two overlapping
broad resonances, see the S-matrix poles in table 1. It should be
detectable although in the same energy region as the $2^{+}$-resonance.
The nuclear $0^{+}$ strength function resembles the plane wave result
reflecting broader underlying structures.

\paragraph*{Conclusions.} We have formulated a method to
compute low-energy three-body continuum spectra for arbitrary
short-range potentials. It is based on a recent successful method used
to calculate bound states by solving the Faddeev equations in
coordinate space.  The angular part of the equations are treated
purely numerically at short distances, whereas the large-distance
behavior of eigenvalues and eigenfunctions is computed essentially
analytically for all partial waves. Combining the results from these
two regions allow accurate computations up to very large distances.
Realistic computations for ground state properties, transition matrix
elements, phase shifts, resonance energies and widths of
$J^{\pi}=0^{\pm},1^{\pm},2^{\pm}$ are carried out for the Borromean
halo nucleus $^6$He. The established $J^{\pi}=2^{+}$ resonance is
found together with a number of other broader resonances.

\paragraph*{\bf Acknowledgments.} One of us A.C. acknowledges the 
support from the European Union through the Human Capital and Mobility
program contract nr. ERBCHBGCT930320.


\begin{thebibliography}{99}

\bibitem{glo96} W.Gl\"{o}ckle, H.Witala,
D.H\"{u}ber, H.Kamada and J.Golak, Phys. Rep. {\bf 274}, 107 (1996).

\bibitem{fri95} J.L.Friar, Proc. Int. Conf. on Few Body Problems in
Physics, Williamsburg 1994, ed. F.Gross, AIP Conference proceedings
{\bf 334}, 323 (1995).

\bibitem{car93} J.~Carbonell, C.~Gignoux, and S.~P.~Merkuriev,
Few-Body Systems {\bf 15}, 15 (1993).

\bibitem{fri95a} J.L. Friar, G.L. Payne, W.Gl\"{o}ckle, D.H\"{u}ber and
H.Witala,  Phys. Rev. {\bf C51}, 2356 (1995).

\bibitem{kie96} A. Kievsky, S. Rosati, W. Tornow and M.Viviani,
Nucl.~Phys.~{\bf A607}, 402 (1996).

\bibitem{han95} P.G.~Hansen, A.S.~Jensen and B.~Jonson,
Ann.~Rev.~Nucl.~Part.~Sci. {\bf 45}, 591 (1995).

\bibitem{fed94} D.V. Fedorov, A.S. Jensen, and K. Riisager, Phys. Rev.
{\bf C49}, 201 (1994); {\bf C50}, 2372 (1994).

\bibitem{zhu93} M.V.~Zhukov, B.V.~Danilin, D.V.~Fedorov, J.M.~Bang,
I.J.~Thompson, and J.S.~Vaagen, Phys. Rep. {\bf 231}, 151 (1993).

\bibitem{efi90} V.N.~Efimov, Comm. Nucl. Part. Phys. {\bf 19}, 271 (1990).

\bibitem{esr96} B.D.~Esry, C.D.~Lin and C.H.~Greene, 
Phys.~Rev.~{\bf A54}, 394 (1996).

\bibitem{dan93a} B.V.~Danilin and M.V.~Zhukov,  Phys. Atom. Nucl.  
 {\bf 56}, 460 (1993).


\bibitem{dan93b} B.V.~Danilin, T.~Rogde, S.N. Ershov,
H.Heiberg-Andersen, J.S. Vaagen, I.J. Thompson and M.V. Zhukov, Phys.
Rev. {\bf C55}, R577 (1997).

\bibitem{fed93} D.~V.~Fedorov and A.~S.~Jensen, Phys.~Rev.~Lett. {\bf
71}, 4103 (1993).

\bibitem{fed96} D.~V.~Fedorov and A.~S.~Jensen, Phys.~Lett. 
{\bf B389}, 631 (1996).

\bibitem{fed94a}D.V.~Fedorov, A.S.~Jensen and K.~Riisager, Phys.~Rev.~Lett.
 {\bf 73}, 2817 (1994). 

\bibitem{tay72} J.R. Taylor, Scattering Theory, (Wiley and Sons, 
New York 1972) Chapter 20.

\bibitem{jen97}  A.S. Jensen, E. Garrido and D.V. Fedorov, Few-Body Systems
(in press). 

\bibitem{gar97}  E. Garrido, D.V. Fedorov and A.S. Jensen, Nucl. Phys.
{\bf A 617}, 153 (1997).

\end{thebibliography}
\end{document}